\documentclass[preprint,5p]{elsarticle}
\usepackage{graphics}
\usepackage{graphicx}
\usepackage{dcolumn} 
\usepackage{bm}
\usepackage{color}
\usepackage{epsfig}
\newcommand{\ccg}{Cr$_2$CoGa}
\newcommand{\cca}{Cr$_2$CoAl}

\begin{document}
\title{Stability of Room Temperature Compensated Half-Metallicity\\
in Cr-based Inverse-Heusler Compounds}
\author[1]{Hyo-Sun Jin}
\author[1,2]{Kwan-Woo Lee\corref{cor1}}
\ead{mckwan@korea.ac.kr}

\cortext[cor1]{corresponding author}
\address[1]{Department of Applied Physics, Graduate School, Korea University, Sejong 30019, Korea}
\address[2]{Division of Display and Semiconductor Physics, Korea University, Sejong 30019, Korea}
\date{\today}

\begin{abstract}
Using three correlated band approaches, namely the conventional band approach plus on-site Coulomb repulsion $U$, 
the modified Becke-Johnson functional, and hybrid functional, we have investigated 
inverse-Heusler ferrimagnets Cr$_2$Co${\cal Z}$ (${\cal Z}$=Al, Ga, In).
These approaches commonly indicate that the Cr$_2$CoAl synthesized recently
is a precise compensated half-metal (CHM), 
whereas Cr$_2$CoGa and Cr$_2$CoIn are ferrimagnets with a small moment.
This is also confirmed by the fixed spin moment approach. 
Analysis of the Bader charge decomposition and the radial charge densities
indicates that this contrast is due to chemical differences among the ${\cal Z}$ ions.
Additionally, in Cr$_2$CoAl, changing the volume by $\pm$ 5\% or the ratio of $c/a$ by $\pm$ 2\% 
does not alter the CHM state, 
suggesting that this state is robust even under application of moderate pressure or strain.
Considering the observed high Curie temperature of 750 K,
our results suggest that Cr$_2$CoAl is a promising candidate for robust high $T_C$ CHMs.
Furthermore, the electronic structure of the CHM Cr$_2$CoAl is discussed.
\end{abstract}
\maketitle

\begin{figure}[tbp]
{\resizebox{7cm}{6.5cm}{\includegraphics{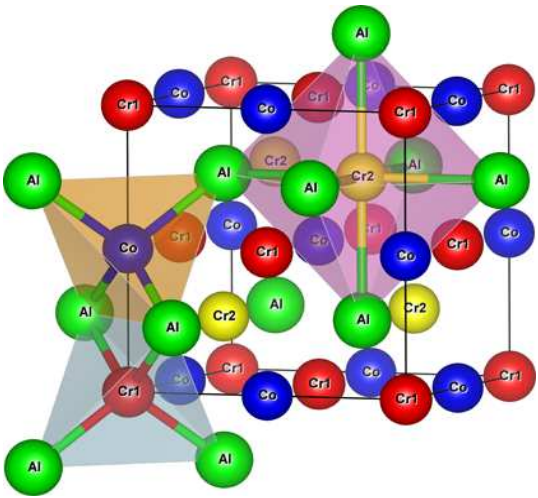}}}
\caption{Inverse Heusler $XA$ (or $X_\alpha$) structure of \cca,
consisting of Cr1-Al$_4$ and Co-Al$_4$ tetrahedra, and Cr2-Al$_6$ octahedra. 
The Cr1-Cr2-Co-Al sequence appears along the diagonal direction.
Here, the two tetrahedra are edge-shared with each other,
and face-shared with the octahedra.
In the ground state, the spin moment of Cr1 ion is antialigned 
to that of both Cr2 and Co ions.
}
\label{str}
\end{figure}

\begin{table*}
\caption{Calculated total and local magnetic moments (in units of $\mu_B$) 
of Cr$_2$YZ (Y= Co, Fe; Z=Al, Ga, In, Ge), 
using GGA, +U, +mBJ, and +EECE approaches.
Here, GGA+U results are given only for $U_{eff}=$ 3 eV for Cr and 4 eV for Fe and Co.
The CHM character remains unchanged even as the strength of $U_{eff}$ 
is varied in the region of 3 -- 5 eV (see text for details). 
Small values of Z ions and interstitial contributions in each compound are not given below.
For the Cr$_2$CoIn unsynthesized yet, our optimized lattice parameter, 
which is similar to the previously reported value,\cite{cca_the} is provided.
}
\begin{tabular}{ccccccccc}\hline\hline
 \multicolumn{1}{c}{Compounds}&\multicolumn{1}{c}{Lattice constants (\AA)}&\multicolumn{2}{c}{Experiments}&\multicolumn{5}{c}{Our calculations}\\\cline{3-4}\cline{5-9}
&&$T_{C}$ (K)&Moments ($\mu_B$)
&Method&\multicolumn{4}{c}{Moments ($\mu_B$)}\\ \cline{6-9}
&&&&&
Cr1 & Cr2 &Y&total\\ \hline
Cr$_2$FeGe&$a=4.629$, & &0.71&GGA&-2.38&1.95&2.08&1.4 \\
&$c=12.414$\cite{hak13}&&&$+U$ &-2.21&1.73&1.7&0.97 \\
&&&&+mBJ&-2.98&2.01&2.21&0.79 \\
&&&&+EECE&-3.04&2.25&3.17&2.09 \\ \hline

Cr$_2$CoGa&$a=5.79$\cite{feng15, deka16}&$\sim$320&0.26 -- 0.31&GGA&-1.9&1.68&0.42&0.07 \\
&&&&$+U$&-1.9&1.15&0.97&0.00 \\
&&&&+mBJ&-2.35&1.68&1.0&0.04\\
&&&&+EECE&-3.02&2.08&1.36&0.06 \\ \cline{2-9}

&$a=5.63$, &440 -- 750&0.21 -- 0.46&GGA&-1.6&1.44&0.34&0.07 \\
&$c=5.88$\cite{jamer16}&&&$+U$&-1.6&0.97&0.82&0.00 \\
&&&&+mBJ&-2.14&1.54&0.9&0.04 \\
&&&&+EECE&-2.87&1.98&1.27&0.04 \\ \hline

Cr$_2$CoIn&$a=5.990$  &--&--&GGA&-2.44&2.14&0.58&0.13 \\
&&&&$+U$&-2.41&1.42&1.22&0.00 \\
&&&&+mBJ&-2.85&2.07&1.21&0.12 \\
&&&&+EECE&-3.34&2.49&1.54&0.32 \\\hline

Cr$_2$CoAl&$a=5.794$\cite{cca_exp}&750&low-moment&GGA&-1.76&1.59&0.33&0.01 \\
&&&&$+U$&-1.75&1.15&0.85&0.00 \\
&&&&+mBJ&-2.12&1.65&0.77&0.00 \\
&&&&+EECE&-2.96&2.16&1.23&0.00 \\
\hline\hline

\end{tabular}
\label{table1}
\end{table*}

\section{Introduction}
The compensated half-metal (CHM), also called the half-metal antiferromagnet,\cite{groot} 
is a half-metal, which has one metallic and the other insulating spin channels, 
with the precisely compensated net moment.
It is expected that this character have many advantages in spintronics applications 
which feature zero stray field and high ordering temperatures.
Although CHMs have been proposed in diverse crystal structures,\cite{nicro}
perovskite- or Heusler-related systems have been regarded
as the most promising candidates.
\cite{nicro,pickett,scoo,pdcro,ku12,felser1}
Recently, Felser and coworkers realized CHM in the Heusler Mn$_3$Ga, 
in which a small band overlap was observed,\cite{felser1} 
by substitution of some Mn ions with Pt ions.\cite{felser2}
Kurt {\it et al.} also observed the CHM in Ru-doped Mn$_2$Ga thin films.\cite{kurt}
However, realization of CHM has been very limited
and the ordering temperature has remained relatively low.

In this paper, we investigated the inverse-Heusler compounds of the type $X_2YZ$ 
with a $X-X-Y-Z$ sequence along the diagonal direction in a cubic
(the full-Heusler compounds have a $X-Y-X-Z$ sequence), 
as visualized in the structure of \cca~ shown in Fig. \ref{str}. 
Using the conventional band theory,
Meinert and Geisler calculated formation energies and magnetic moments of
several inverse-Heusler systems, 
and predicted the formation of ferrimagnetic states with a tiny moment.\cite{cca_the}
Based on this theoretical suggestion, 
a few compounds including Cr$_2$CoGa and \cca~
have been synthesized and experimentally investigated. 
Jamer {\it et al.} recently synthesized \cca~ and observed 
a high Curie temperature $T_C\approx750$ K,\cite{cca_exp}
which was much higher than that in the CHM Mn$_{3-x}$Pt$_x$Ga 
whose $T_C$ was below 300 K.\cite{felser2}
The x-ray magnetic circular dichroism measurements indicated
a tiny net moment resulting from an antialigned configuration of Cr and Co moments,
but due to some impurity phases in the sample
it remained unclear whether the net moment was tiny or exactly zero.
Additionally, several research groups have observed that the magnetic moment in Cr$_2$CoGa, 
a promising CHM candidate with high $T_C$, 
is very sensitive to the synthesizing conditions.\cite{hak13,feng15,deka16,jamer16}
It may be due to antisite disorder, which often occurs in the Heusler compounds.
These findings necessitate a fine {\it ab initio} calculation, 
including correlation effects not considered previously 
but which are often crucial in transition metal compounds,
to clarify the magnetic ground states.
Besides, for further experimental researches it will be useful to investigate 
how the CHM state is robust under external conditions such as strain and pressure.

Here, we performed fine {\it ab initio} calculations with very dense $k$-meshes,
using correlated band theories of the generalized gradient approximation 
plus $U$ (GGA+U),\cite{amf} the Tran and Blaha modified Becke-Johnson functional (mBJ),
\cite {mbj,comm}and the hybrid functional\cite{hybrid}.
Our results indicate that \cca~ is the most promising candidate
of an exact CHM, which is also affirmed by the fixed spin moment approach.\cite{fsm} 
We also investigate the stability of this state under pressure or strain.
Moreover, we address the electronic structure of the CHM state, 
which is not reported till now.

\section{Structure and calculation}
In the cubic inverse Heusler $X_2YZ$ structure (space group: $F\bar{4}3m$, No. 216), 
two types of $X$ ions lie at $4a$ (0,0,0)
and $4c$ ($\frac{1}{4},\frac{1}{4},\frac{1}{4}$).
The $Y$ and $Z$ ions lie at $4b$ ($\frac{1}{2},\frac{1}{2},\frac{1}{2}$) and 
 $4d$ ($\frac{3}{4},\frac{3}{4},\frac{3}{4}$), respectively.
Some of the inverse Heusler compounds, such as Cr$_2$FeGe and Cr$_2$CoGa, 
show a tetragonal distortion.
In the tetragonal structure (space group: $I\bar{4}m2$, No. 119),
$X$ ions lie at $2b$ (0, 0, $\frac{1}{2}$) and $2d$ (0, $\frac{1}{2}$, $\frac{3}{4}$)
and the $Y$ and $Z$ ions at $2a$ (0, 0, 0) and $2c$ (0, $\frac{1}{2}$, $\frac{1}{4}$), respectively.  
The lattice parameters used in this study are given in Table \ref{table1}.
For all compounds, experimentally obtained values were used in our calculations,
except for Cr$_2$CoIn which has not been synthesized yet.
Note that our optimized parameters were found to be within 3.5\% of the experimental values,
so that very little deviation occurred in the magnetic properties. 

Our calculations were carried out with the Perdew-Burke-Ernzerhof GGA 
as the exchange-correlation functional,
\cite{gga} implemented in the accurate all-electron full-potential code {\sc wien2k}\cite{wien2k,dft}
Correlation effects in the GGA+U approach were treated 
by the around mean field double-counting scheme,\cite{amf}
which is often proper for the moderate correlated systems 
such as the present systems.\cite{LP_nacoo}
In these calculations, we varied the effective on-site Coulomb repulsion parameter $U_{eff}=U-J$
in the range of $3-5$ eV for the Cr, Fe, and Co ions, as commonly applied.\cite{pdcro,srcro,sfro,lcmo,lco} 
Here, $J$ is an intra exchange integral.
Additionally, two other correlated band approaches of the hybrid functional\cite{hybrid} 
and the mBJ\cite{mbj}, implemented in {\sc wien2k}, were used to confirm our GGA+U results.
In the hybrid functional approach, the exact exchange for correlated electrons functional (EECE) 
for treating correlated electron systems was chosen with a common 25\% of local density
exchange replaced by the exact on-site exchange.\cite{ucd16}
All results obtained here are very consistent, 
although there are subtle differences in some cases.
As expected, spin-orbit coupling was excluded, 
since our preliminary calculations showed negligible effects due to this in these systems. 

In the {\sc wien2k}, the basis was determined by $R_{mt}K_{max}=7$ 
and the augmented atomic radii of 2.28 -- 2.5 a.u. for $X$ and $Y$ ions and 2.18 -- 2.39 a.u. for $Z$ ions.
A very dense $k$-mesh containing up to $30\times30\times30$ points was used
to check the convergence of the tiny moment character 
as a more careful treatment was required near the Fermi level $E_F$.

\section{Results}

\subsection{Magnetic states in the correlated region}
Recently, several ferrimagnetic (FI), transition metal-based inverse-Heusler compounds 
with low moments have been synthesized or theoretically proposed.
However, as mentioned in the section of Introduction, 
the low moment character remains ambiguous.
In the experimental point of view,
the observed moments are very sensitive to the experimental conditions.
In the theoretical point of view,
the correlation effects usually relevant to transition metal based compounds 
have been excluded in the existing calculations.
We now investigate these compounds through fine calculations, 
using the three correlated band approaches based on the density functional theory,
with a very dense $k$-mesh. The results are given in Table \ref{table1}.

We first address the tetragonal Cr$_2$FeGe with the observed net moment of 0.71 $\mu_B$,\cite{hak13}
as a test system to check whether our approaches are reasonable for the Cr-based compounds.
The spin of Cr1 ions is antialigned to both the spins of Cr2 and Fe ions,
but the moment obtained from GGA is larger than the observed value by a factor of 2.
On the other hand, inclusion of correlation using GGA+U and +mBJ approaches 
yields very similar moments in the range of 0.79 -- 0.97 $\mu_B$ 
as those obtained from the experiment.
As expected, these results affirm the crucial role of correlation in such Cr-based compounds.
Note that the GGA+EECE shows a substantially enhanced moment in Cr$_2$FeGe, 
whereas in the other compounds studied here the results of all three approaches are very similar.

Next, we investigate the FI \ccg, which is experimentally the most well-studied among all three so far. 
\ccg~ has both cubic and tetragonal phases with similar magnetic moments.\cite{feng15, deka16,jamer16}
However, the observed Curie temperature $T_C$ (in the range of 440 -- 750 K) 
and moment (0.21 -- 0.46 $\mu_B$) are sensitive to the synthesizing conditions.\cite{jamer16}
Consistently with the experiments, 
in all our calculations the spin of Cr1 ions is antialigned to both the spins of Cr2 and Co ions,
leading to a tiny net moment, but is not exactly compensated.
The small moment of \ccg~ is negligibly affected by the correlation.
So, both experiments and calculations consistently indicate that \ccg~ is not an exact CHM.

Further, the magnetic properties of the two isovalent and isostructural compounds 
namely, cubic Cr$_2$CoIn and \cca~ are evaluated. 
For Cr$_2$CoIn, which has not been synthesized so far, 
only GGA+U produces a precise CHM state, 
whereas all other results show a small net moment of 0.12 -- 0.32 $\mu_B$.
In the case of \cca, in the GGA calculations 
a Fermi energy $E_F$-crossing bands around the $\Gamma$-point appear in the spin-down channel (see below). 
Consequently, a small moment of 0.01 $\mu_B$ results 
which is consistent with the findings of an earlier study.\cite{cca_the} 
Considering correlation effects, 
the bands are completely filled, resulting in half-metallic (see below).
In this state, the net moment of \cca~ is exactly compensated.
The origin of this distinction among these compounds will be identified below.
Note that triple nodal points near the Fermi energy in the spin-up lead to a nontrivial topological feature, 
as will be addressed elsewhere.\cite{nexus}

\begin{figure}[tbp]
{\resizebox{8cm}{6cm}{\includegraphics{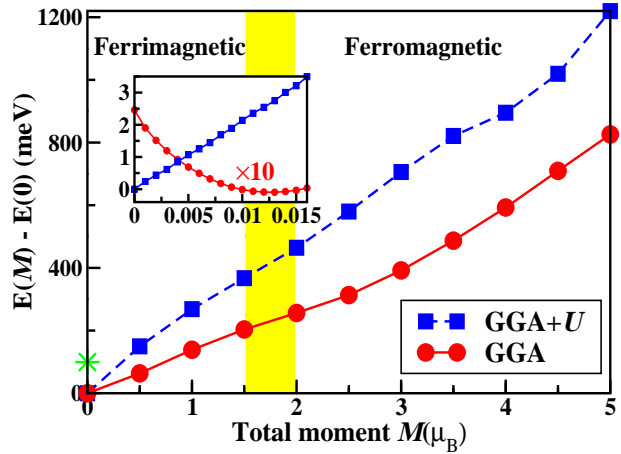}}}
\caption{For \cca, change in energy versus fixed spin moment $M$ 
(in units of $\mu_B$ per formula unit) in GGA and GGA+U.
Ferrimagnetic state (FI) is favored over ferromagnetic state (FM) for $M \le 1.5 \mu_B$, 
while FM dominates FI for $M \ge 2 \mu_B$. 
In the shaded region of $1.5 < M < 2.0$, both states coexist.
The symbol $\ast$ denotes the energy of the nonmagnetic state.
Here $U_{eff}=3$ eV for Cr and 4 eV for Co are used.
$E(0)$ is the energy of the CHM state.
Inset: Blowup plots below $M\sim0.2 \mu_B$,
indicating that CHM is the ground state in GGA+U.
In the Inset, the energies of GGA are enhanced by a factor of 10 for a better visualization.  
}
\label{fsm}
\end{figure}

\subsection{Studies of fixed spin moment of Cr$_2$CoAl}
We performed fixed spin moment (FSM) calculations to confirm the CHM character of \cca.
Figure \ref{fsm} shows the plot of energy increment $\Delta E=E(M)-E(0)$, 
relative to the energy $E(0)$ of the CHM state, 
versus fixed spin moment $M$ in both GGA and GGA+U methods. 
The nonmagnetic state has a higher energy by approximately 100 meV 
than the CHM energy $E(0)$ as obtained from both GGA and GGA+U.   
Except for the low moment region, both the GGA and GGA+U results 
show a similar behavior in which the energy increases monotonically with increase in $M$.
The inset of Fig. \ref{fsm} shows a magnified view of the low moment region 
where the ground state in GGA appears at $M\approx0.012 \mu_B$,
consistent with the self-consistent results.
However, in the GGA+U, at the region the energy difference $\Delta E$ linearly increases 
from the value at $M=0$.
The linear behavior of the energy increment 
is testimony to the character of CHMs or half-metals.\cite{scoo}

\begin{figure}[tbp]
{\resizebox{7.6cm}{5.8cm}{\includegraphics{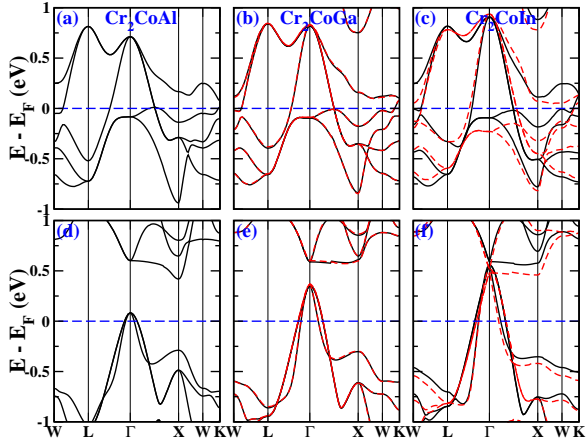}}}
\caption{Comparison of band structures in GGA.
Top and bottom lines are for the majority and minority spin channels, respectively.
The (black) solid lines are for the original structures,
and the (red) dashed lines are for the identical structure with \cca.
Since difference in structure between \cca~ and \ccg~ is small, 
the change in structure leads to no distinguishable variation, as shown in (b) and (e).
Even for the identical structure, substantial differences among these three systems are clearly visible 
in the spin-down channels.
}
\label{bs2}
\end{figure}

\begin{figure}[tbp]
{\resizebox{7.6cm}{5.8cm}{\includegraphics{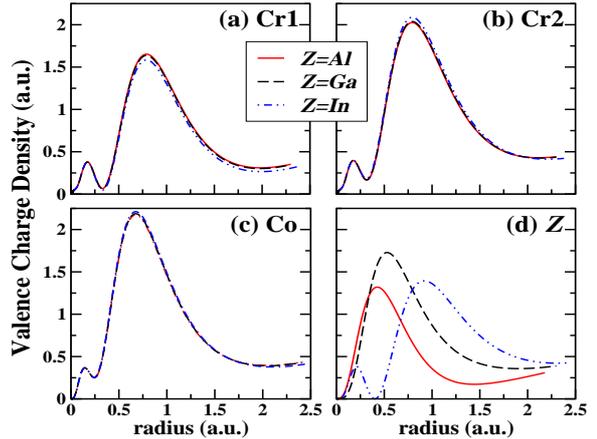}}}
\caption{Radial charge densities $4\pi r^2\rho(r)$ of each ion in Cr$_2$Co${\cal Z}$ compounds
versus distance from the each ion nucleus, in a.u.
The significant difference in the charge distribution among the ${\cal Z}$ ions
substantially varies that of that Cr1 and Cr2 ions, 
while that of the Co ion is affected somewhat.
}
\label{radchar}
\end{figure}

\begin{table}
\caption{Bader charges of Cr$_2$Co${\cal Z}$ (${\cal Z}$=Al, Ga, In) 
obtained from GGA and GGA+U.
}
\begin{tabular}{cccccc}\hline\hline
 \multicolumn{2}{c}{}&\multicolumn{4}{c}{Charges}\\\cline{3-6}
 Method&~& Cr1 & Cr2 & Co & ${\cal Z}$ \\ \hline
 ~ & Al  & +0.37 & --0.28 & +1.33 & --1.42 \\
GGA & Ga & --0.29 & --0.39 & +0.52 & +0.16 \\
~    & In & --0.28 & --0.42 & +0.57 & +0.14 \\\hline
 ~ & Al  & +0.40 & --0.22 & +1.23 & --1.41 \\
+U & Ga & --0.36 & --0.25 & +0.36 & +0.25 \\
~    & In & --0.37 & --0.27 & +0.37 & +0.25 \\
\hline\hline
\end{tabular}
\label{table2}
\end{table}

\subsection{Effects of pressure and strain on magnetic state}
To investigate the impact of pressure and the stability of the CHM state in \cca, 
we first vary the volume by $\pm$5\% of the experimental value.
Remarkably, the net moment remains unchanged in all correlated calculations within this range of volume,
except in the EECE case which showed a tiny moment of less than few hundredths of a $\mu_B$ 
when the volume was enhanced by more than 3\%.
Second, the ratio of in-plane and out-of-plane lattice constants is varied to inspect
the effect of strain.
A small strain does not lead to changing the moment, 
but a larger strain by over $\pm$2\% induces a tiny moment of about a few thousandths $\mu_B$.
This is consistent with the tiny difference in the magnetic moments between the cubic and tetragonal Cr$_2$CoGa.
It is worthy to be noted that the magnetic moment of Cr$_2$CoIn also remains nearly unchanged,
when compressing the volume by 5\% or applying a strain.
These results indicate that the magnetic states are very robust under application of pressure or strain.

\subsection{Identifying differences}
These differences among the three Cr-based inverse-Heusler compounds
can be clarified in two ways.
The difference in ionic sizes of ${\cal Z}$ atoms leads to structural differences.  
We first compared band structures with the same structure of \cca~ to show effects
of pure chemical difference.
As shown in Fig. \ref{bs2}, even in the identical structure 
distinctions among band structures of these three systems are very evident near $E_F$,
indicating small effects of structural difference.
This suggests that applying pressure is not a useful tool to achieve a CHM state
in the system, as mentioned above.

Next, we calculated the Bader charges from the {\sc wien2k} code, given in Table \ref{table2}.
Remarkably, the Al ion is much more ionic than the Ga and In ions.
This is also observed in the radial charge density, which is  a more efficient way 
to analyze differences in ionic charges.\cite{radchar12}
As shown in Fig. \ref{radchar},
the charge of the Al ion is distributed more closely to the nucleus than that of the other ${\cal Z}$ ions.
This difference in the charge distribution of ${\cal Z}$ ions affects 
the charge distribution of Cr1 and Cr2 ions,
resulting in considerable variations of their local magnetic moments in each compound.
(See Table 1.)
These results confirm that the chemical differences among ${\cal Z}$ atoms, rather than the structure, 
lead to distinctions in the magnetic properties.
One may expect to achieve an exact CHM state in Cr$_2$CoGa and Cr$_2$CoIn by some Al-doping,
so requiring more research.

\begin{figure}[tbp]
{\resizebox{8cm}{6cm}{\includegraphics{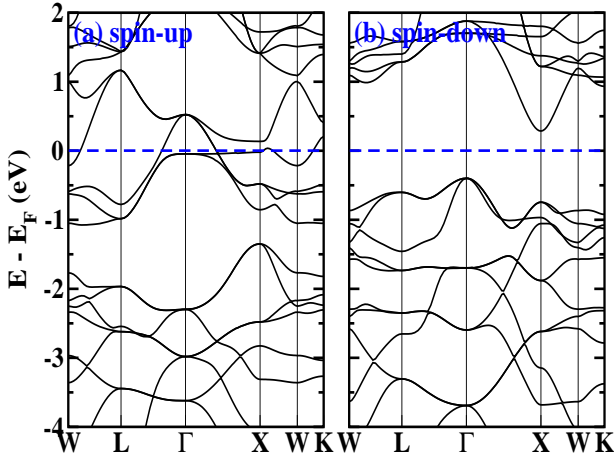}}}
\caption{(a), (b) Spin-resolved GGA+U band structures, showing a half-metallic character.
Here $U_{eff}=3$ eV for Cr and 4 eV for Co are used.
The horizontal dashed line indicates the Fermi level $E_F$, which is set to zero.
}
\label{band}
\end{figure}

\subsection{Electronic structures of Cr$_2$CoAl}

Finally, we address the electronic structures of the CHM \cca~ in the correlated regime,
which were not reported before.
There are some distinguishing features of the band structures obtained here 
by the three different correlated band approaches,
but they are very similar to each other as a whole. 
Therefore, only the GGA+U results are presented here, unless mentioned otherwise.

Consistent with the FSM results, the CHM state is energetically favored 
over the nonmagnetic state by 100 meV per formula unit.
Our trials to obtain a ferromagnetic state are always converged into a FI state, 
indicating that the latter is much more stable.

\begin{figure}[tbp]
{\resizebox{8cm}{6cm}{\includegraphics{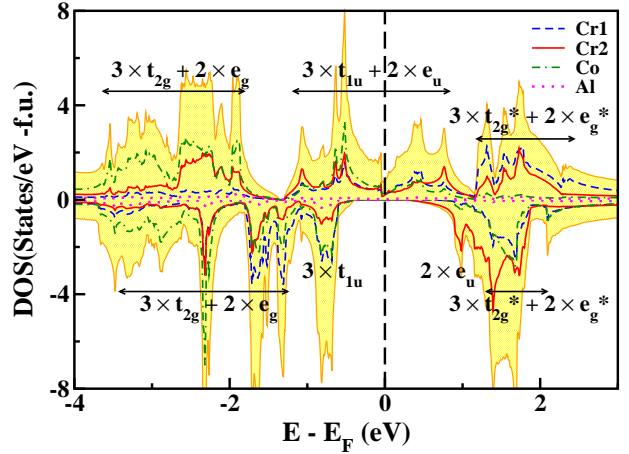}}}
\caption{Total and atom-resolved densities of states (DOSs)
in GGA+U with $U_{eff}=3$ eV for Cr and 4 eV for Co.
The symbols represent the hybridization feature.
The total DOS, (yellow) shaded region, $N(E_F)$ at $E_F$ 
is 0.047 states/eV per formula unit.
}
\label{dos}
\end{figure}

\begin{figure}[tbp]
{\resizebox{8cm}{4.0cm}{\includegraphics{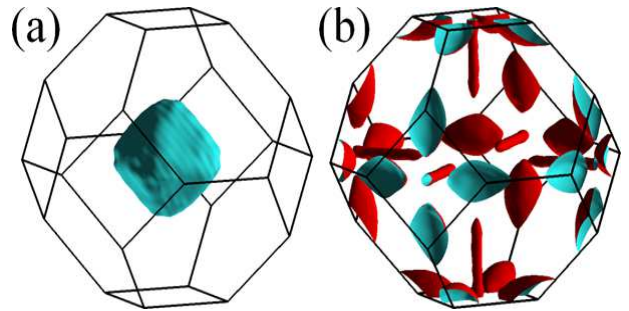}}}
\caption{GGA+U Fermi surfaces, 
which possess (a) holes and (b) electrons
}
\label{fs}
\end{figure}

The spin-resolved band structures of the GGA+U are illustrated in Figs. \ref{band} (a) and (b).
As depicted in Fig. \ref{dos},
\cca~ shows a hybridization feature very similar to what has been earlier discussed
in other inverse-Heusler compounds.\cite{gal}
The $d$ orbitals of Cr1-Co ions 
lead to \{$t_{2g}$, $e_{g}$\} manifolds and nonbonding \{$t_{1u}$, $e_{u}$\} orbitals.
These 10 orbitals are hybridized with the Cr2 $d$ orbitals, 
resulting in bonding \{$t_{2g}, e_g$\}, antibonding \{$t_{2g}^\ast, e_g^\ast$\},
and nonbonding \{$t_{1u}$, $e_{u}$\} manifolds.
In the minority channel, an indirect gap of 1 eV appears between the $t_{1u}$ and $e_u$ manifolds.
In the majority channel, a mixture of the $t_{1u}$ and $e_u$, separated from the other orbitals,
is partially filled, indicating a half-metallic character.
Notably, in the metallic channel, 
a nearly flat band appears to cross over $E_F$ along the $\Gamma-X$ line,
as often observed in perovksites without $dd\delta$ hopping.\cite{srcro}
This band leads to two interesting features in the density of states (DOS) and fermiology.
As shown in the GGA+U total and atom-resolved DOSs of Fig. \ref{dos}, 
$E_F$ pinpoints a pseudogap, resulting in a tiny DOS at $E_F$.
As picturized in Fig. \ref{fs} (b), rod-like hole Fermi surfaces (FSs) appear along the $X$ to $\Gamma$ line. 
Additionally, there are $\Gamma$-centered hole and $W$-centered elliptical electron FSs,
given in Fig. \ref{fs}.
The presence of the pseudogap close to $E_F$ in the metallic channel
may lead to unusual transport phenomena like ultrafast demagnetization.\cite{mann}

Although the topological properties are not covered here,
it is worthy to be noted that 
the metallic channel shows remarkable features, seemingly related with topological characters, 
near the Fermi energy $E_F$.
As in the full Heusler Co$_2$MnGa,\cite{hopf} 
a crossing between the nearly flat band and two-fold bands 
appears just below $E_F$ along the $\Gamma-X$ line. (See Fig. \ref{band}(a).)
In the Co$_2$MnGa, four-bands involved this crossing lead to a hopf-link semimetallic state.
Instead, in the inverse Heusler compounds, this crossing leads to six triple points 
due to lack of inversion symmetry along the $\langle 100\rangle$ in the BZ,
implying another interesting topological character.
Recently, Shi {\it et al.} also suggested a Weyl semimetallic state
in the inverse Heusler FI Ti$_2$MnAl.\cite{ifw18}
These interesting topological properties will be discussed elsewhere.\cite{nexus}

\section{Summary}
We have investigated a few Cr-based inverse-Heusler compounds 
Cr$_2$Co${\cal Z}$ (${\cal Z}$=Al, Ga, In)
through fine electronic structure calculations based on 
three correlated band approaches, which have been used widely,
in the hope of finding promising candidates for compensated half-metals (CHMs)

In these systems, our results show energetically most favored ferrimagnetic state,
in which the spin of one of the two Cr ions is antialigned to that of the other Cr and Co ions.
In contrast to the other systems showing small moments,
the net moment of \cca~ is precisely compensated.
This is also confirmed by the fixed spin moment approach.
Through analysis of the radial charge distribution,
we show that this distinction is due to chemical differences among the ${\cal Z}$ ions,
implying that a precise CHM state would be also obtained by substitution of Al in Cr$_2$CoGa and Cr$_2$CoIn.
Our results also suggest that the CHM state of \cca~ is robust under pressure or strain.
Therefore, considering the observed high $T_C$, 
\cca~ would be a promising candidate for robust compensated half-metals with a high T$_C$, 
requiring more experimental research.

\section{Acknowledgments}
 This research was supported by NRF of Korea Grant No. NRF-2016R1A2B4009579.


\begin{thebibliography}{10}
\bibitem{comm} In the +mBJ calculations, 
the {\sc wien2k} code use only the local spin density approximation (LSDA),
regardless of a type of initial exchange-correlation functionals.
Thus, this is the LSDA+mBJ approach.
\end{thebibliography}
\end{document}